\documentstyle[amssymb,aps,epsf,prl,multicol]{revtex}

\begin{document}
\title{{\large Laser Cooling Concept for Translational Motion of Neutral Molecules }}
\author{C. H. Raymond Ooi\thanks{{\footnotesize ooi@spock.physik.uni-konstanz.de}},
Karl-Peter Marzlin\thanks{{\footnotesize Peter.Marzlin@uni-konstanz.de}} and
J\"{u}rgen Audretsch\thanks{{\footnotesize Juergen.Audretsch@uni-konstanz.de}%
}}
\address{{\it Fachbereich Physik der Universit\"{a}t Konstanz, Fach M674}, D-78457 \\
{\it Konstanz, Germany}}
\date{\today}
\maketitle

\begin{abstract}
We propose a laser cooling concept for translational motion of neutral
molecules. The cooling works by repetition of a cooling cycle composed of
three sequential steps: velocity selection of a narrow momentum width,
deceleration of velocity selected molecules and irreversible accumulation.
Those molecules which have been cooled undergo only one optical spontaneous
emission for irreversible accumulation. This circumvents the problem of
broken closed pumping cycle required for laser cooling by optical
spontaneous emissions. We consider thermalized molecules at temperature of $%
1K$ as precursor and describe a rotational pumping scheme to prepare the
molecules in an appropriate initial states for translational cooling. Using
realistic cooling parameters, we estimate the cooling time and show that
efficient cooling to a momentum width of $\sim \hbar k$ can be achieved
within a practical timescale. The physical feasibility of the cooling
concept is showned by developing a cooling model. This enables analytical
description of the transient populations and entropies. Simulation shows
that after the cooling process, a large momentum width is reduced to a
narrow final width, determined by velocity selection. Translational cooling
occurs through the reduction of the center of mass entropy, mainly during
coherent laser interactions. The cooling process removes the translational
entropy irreversibly to the radiation and internal entropy sinks.
\end{abstract}

\section{Introduction}

The existing laser cooling schemes for atoms have been successfully based on
the concept of repeated optical excitation followed by spontaneous emission 
\cite{atomcool}. In molecules, there are additional motional degree of
freedoms, namely the vibrational and rotational motions in addition to the
center of mass translational motion. This creates two characteristics in
molecules which are absent in atoms. First, the optical spontaneous emission
(fluorescence) will inevitably repopulate more than one state. This makes
existing cooling schemes for atoms unapplicable to molecules. Second, the
existence of additional quantum states associated with the vibrational and
rotational motions provide additional entropy sinks other than the radiation
reservoir for translational cooling.

A scheme for sequential translational-vibrational-rotational cooling \cite
{stwalley96} has been proposed but not demonstrated yet. Existing schemes
for sole vibrational cooling \cite{vibcool} use optimally controlled
ultrashort pulses. A translational cooling scheme which uses cavity for
incoherent processes of dissipation has been proposed \cite{cavitycool}.
Laser cooling of translational motion of molecules is more challenging due
to the requirement of large momentum kick from optical photons to slow a
molecule and the difficulty of maintaining an efficient closed optical
pumping cycle, since fluorescence from the excited electronic state will
populate other molecular levels and leads to inefficient returning to the
original ground state. Closed cycles in far-infrared and microwave regimes
are available but may not be favorable due to their low spontaneous emission
rates. Thus, a closed optical pumping cycle which involves repeated{\em \
optical }spontaneous emissions cannot be simply established in molecules for
continuous removal of the translational kinetic energy and entropy.

In this work, we circumvent the problem by first establishing efficient
closed optical pumping cycles based on coherent processes of Raman velocity
selection and STIRAP (Stimulated Raman Adiabatic Passage)\cite{gaubatz} to
continuously reduce the translational kinetic energy of a narrow velocity
class of molecules to near zero. Then, this cooled class of molecules is
pumped to irreversibly populate other rovibrational states through a single
optical spontaneous emission. These molecules need no further recycling and
are decoupled from lasers. This is the main ingredient for successful
translational cooling of molecules. Our cooling scheme do not introduce
permanent vibrational heating, since the molecules subsequently undergo
infrared decay mainly to the low vibrational levels in about $1s$. As in
existing vibrational cooling schemes \cite{vibcool}, the decay inevitably
leads to some rotational heating. If necessary, this would need a separately
rotational cooling scheme, which will be reported in our future work. Our
aim here is for translationally cooled molecules, which would be invaluable
assets for molecular physics since they can be confined in shallower and
more handy traps for convenient manipulations, fundamental studies and
further internal coolings.

In Section II, we elaborate our cooling concept in a general way. Then, in
Section III we describe how the molecules to be translationally cooled, can
be appropriately prepared in practice using a rotational pumping scheme and
buffer gas cooled molecules\cite{buffer}. The cooling scheme is shown to be
feasible in two aspects; in practice (as shown in Section IV by estimates of
physical quantities) and in principle (as shown in Section V by modelling of
the cooling process). We show that the velocity selection, normally used for
subrecoil cooling \cite{Aspect}\cite{Kasevich} can be practically used to
cool molecules from $1K$\ to Doppler temperature. We model the cooling
process with analytical density matrix and study the entropy change of the
system. The translational entropy is found to reduce mostly during the
coherent deceleration and velocity selection steps. Although an irreversible
process from\ the single spontaneous emission per molecule is necessary for
cooling and leads to the increase in the radiation entropy, it does not
contribute substantially to the translational entropy reduction.

\section{Cooling Concept}

In this paper, we propose a laser cooling concept for the translational
motion of neutral molecules. We describe the cooling process with an
analytical model and demonstrate the cooling results by numerical
simulations. The cooling process is based on the repetition of a cooling
cycle, each composed of three different physical steps: 1) velocity
selection, 2)deceleration and 3)accumulation. We refer to the molecules as a%
{\it \ }{\em total system }(subscript `tot'){\em \ }with two subsystems: the
electronic, vibrational and rotational states as the {\em internal }degree
of freedom (subscript `I'){\em \ }and the molecular center of mass as the 
{\em external} degree of freedom (subscript `cm'). The velocity selection
and deceleration involve 4 internal levels, 3 ground states $|g0,\pm \rangle 
$ corresponding to the states with magnetic quantum number $M=0,\pm 1$ and
an excited state $|e\rangle (M=0)$(Fig. 2a and 2b). {\bf \ }The states $%
|e\rangle ,|g0\rangle $ and $|g+\rangle $ are used for velocity selection
while the deceleration are based on the states $|e\rangle ,|g-\rangle $ and $%
|g+\rangle $. For convenience, we may refer to $|g0\rangle $ and $|g\pm
\rangle $ as {\em initial state} and {\em Raman states} respectively.

The translational cooling starts with the{\em \ first step}, which is the
velocity selection. (see Fig. 2a and 3a). A fraction of \ molecules with
narrow momentum width and mean momentum $P$ is selected from the population
in state $|g0\rangle $ and transfered to an empty state $|g+\rangle $ using
a $\pi $-pulse, without populating the excited state. This step is
essentially the same as described in Ref.\cite{Kasevich} except that here we
use orthogonal $lin-\sigma ^{+}$ lasers instead of counter propagating $%
\sigma ^{+}-$ $\sigma ^{-}$ lasers, since the dipole moment between states $%
|g0\rangle $ and $|e\rangle $ is aligned along the $\sigma ^{+}$ polarized
laser axis.

In the {\em second step }(see Fig. 2b and 3b), the selected molecules are
decelerated using a coherent process. In principle, non-optical deceleration
technique can be applied \cite{dece}. However, for practical reasons, we
require efficient and fast optical momentum transfer by repeated population
inversion between the Raman states $|g+\rangle $ and $|g-\rangle .$ This can
be accomplished by various population transfer techniques using optimally
engineered laser pulses \cite{advpoptrf}. Here, we adopt the STIRAP process
which has been shown to be highly efficient in molecules \cite{gaubatz}\cite
{efficient} and has been used for coherent momentum transfer in atom optics
and atom interferometry \cite{Marte}. In this step, the internal populations
are repeatedly inverted between $|g+\rangle $ and $|g-\rangle .$ Each
inversion is a {\em substep} of step 2, which is accomplished by reversing
the directions and the counter-intuitive sequence of a pair of counter
propagating $\sigma +$ and $\sigma -$ polarized pulses. Each pair of the
STIRAP pulse provides about $2\hbar k$ of photon momentum kick to decelerate
the molecules. Thus, the population with a mean momentum $P$ will require a
number of $P/2\hbar k$ STIRAP inversions to reduce the mean momentum $P$
towards zero. In the end of this step, we have a fraction of molecules with
the translational energy close to zero .

We want to repeat steps 1 and 2 to accumulate the populations with narrow
momentum width around zero momentum. Since steps 1 and 2 are coherent
processes, subsequent repetition of these steps would not lead to net
cooling because the next STIRAP process would accelerate the molecules out
of zero mean momentum. The Raman states must be emptied before the next
deceleration cycle starts. We achieve this with an irreversible process in
the {\em third step}(see Fig. 2c and 3c){\em .} The cooled and decelerated
molecules are rapidly transferred from the Raman states to a decaying state $%
|d\rangle $ which may be an excited electronic state or a rovibrational
state in the ground electronic state. Due to the requirement of fast
translational cooling of unconfined molecules, we use an excited electronic
state which provides fast optical spontaneous decay time. Subsequent
spontaneous emissions populate the many vibronic ground as the accumulation
states $\{|acc\rangle \}.$ We emphasize that there is {\em no} vibrational
heating as time passes, since the molecules in states $\{|acc\rangle \}$ are
decoupled from the laser beams and will undergo tens of infrared spontaneous
decays to the lowest vibrational levels in about $1s$. This introduces
additional momentum spread in the order of an optical photon momentum, much
smaller than the overall reduction of momentum width. It is the
irreversibility of the spontaneous emission which enables repeated
accumulation of the molecules with narrow momentum slices and zero mean
momentum from each cooling cycle as steps 1, 2 and 3 are repeated.

Each {\em cooling cycle} which is composed of a sequence of the three steps
described above, begins with the velocity selection of a new slice of narrow
momentum width. The whole cooling process involves the repetition of the
cooling cycle. We point out that the equivalent temperature of the molecules
after the process corresponds to the velocity selected momentum width, which
can be well below the recoil limit.

The main difficulty to laser cool molecules by repeated spontaneous
emissions in the optical regime is due to the fluorescence decay from an
excited electronic state to many rotational-vibrational states of the ground
electronic state. After a spontaneous emission, the molecules have a high
probability of not returning to the original state. Thus, the molecules
decay to a state which is off-resonant with the pumping laser before their
momenta are reduced by the laser photons. In our cooling scheme, the
momentum and kinetic energy of each molecule are quickly reduced toward zero
by deceleration (step 2) {\em before} undergoing a {\em single optical
spontaneous emission} process in the end of a cooling cycle. Thus, each
cooling cycle completely cools a particular fraction of molecules. This
cooling scheme circumvents the requirement of repeated spontaneous emissions
for each molecule to obtain cooling. It is this unique feature which makes
translational laser cooling of molecules possible.

\section{\protect\bigskip Preparation of Initial Molecules}

\ Initially, the molecules to be translationally cooled can be prepared in a
thermal equilibrium of $1K$\ by the versatile buffer gas cooling \cite
{buffer}, with a single heavily populated internal state $|g0\rangle $\ and
two empty states $|g\pm \rangle $\ using a rotational pumping scheme
described below.

\subsection{Thermalized molecules}

We start with the preparation of thermalized molecules which can have
sufficiently few internal states for convenient rotational pumping and
sufficiently low translational temperature for practical translational
cooling in unconfined space. We emphasize that we are dealing with {\em %
neutral} molecules instead of molecular ions. Neutral molecules produced by
laser ablationed are highly excited internally and translationally. Buffer
gas cooling is a good technique to thermalize the molecules, reduce the
translational temperature of $1K$ and depopulate many of the highly excited
vibronic and rotatinal states. Elastic collisions during buffer gas cooling
provides a good thermalization mechanism to bring the molecules close to
thermal equilibrium. At thermal equilibrium of temperature $T_{i}$, the
populations of the ground electronic surface $|\Sigma \rangle $ are
distributed among rovibrational states $(n,J)$ according to the
Maxwell-Boltzmann distribution \cite{mw spect}

\begin{equation}
P(n,J)\propto (2J+1)\exp \{-hcBJ(J+1)-hcv(n+\frac{1}{2})\}/k_{B}T_{i}
\end{equation}
where $v$ and $B$ are vibrational and rotational constants (in wavenumber)
respectively. The maximum population corresponds to $n_{\max }=0,J_{\max
}\approx \sqrt{\frac{k_{B}T_{i}}{2hcB}}-\frac{1}{2}.$ At initial temperature
of $T_{i}=1K$ and $B\approx 0.1cm^{-1},$ the molecules are mainly in the
ground electronic state with the highest populations in $n=0$ and $J_{\max
}\approx 1$ or $2.$ Thus, the molecules are reasonably cold internally and
this makes it easy to create a single heavily populated state $|g0\rangle $
for subsequent translational cooling. Since the excited vibrational states
are scarcely populated, molecular collisions rates are small and can be
neglected \cite{no-collision}.

\subsection{Rotational Pumping}

We wish to create 2 empty states $|g\pm \rangle $ and a heavily populated
state $|g0\rangle $ for more efficient translational cooling. This can be
done by pumping the populations from the most highly populated rotational
sublevels into a single sublevel within the ground vibrational and ground
electronic state, namely $|g0\rangle =|\Sigma ,n=0,J=1,M_{J}=0\rangle $(see
Fig. 1), Using $\sigma ^{+}$ polarized laser, the populations in
rovibrational states $|n,J,\{M_{J}\}\rangle =|0,1,-1\rangle ,$ $%
|0,2,\{-2,-1,0\}\rangle $ are excited to states $|1,0\rangle ,$ $%
|1,1,\{-1,0,1\}\rangle $ \ respectively. On the other hand, $\sigma ^{-}$
polarized laser (same frequency as $\sigma ^{+}$) excites.the populations
from states $|0,1,1\rangle ,$ $|0,2,\{2,1,0\}\rangle $ to states $%
|1,0\rangle ,$ $|1,1,\{1,0,-1\}\rangle $ respectively. A third laser, linear
polarized with higher frequency excites the population in $|0,0\rangle $ to $%
|1,1,0\rangle .$ Subsequent infrared decays by virtue of selection rules, $%
\Delta J=\pm 1$ and $\Delta M_{J}=0,\pm 1$ effectively transfer the
populations from the above states to a single state $|0,1,0\rangle $%
(referred as $|g0\rangle $). The empty states of $|1,1,\pm 1\rangle $ are
referred as $|g\pm \rangle .$ The $\sigma ^{+}$ and $\sigma ^{-}$ polarized
lasers are switched on alternatively instead of simultaneously to avoid the
creation of a dark state which will prevent complete pumping to state $%
|g0\rangle .$ Taking into account the magnetic levels splitting by the
trapping field and small differences in the rovibrational transition
frequencies, we could use the lasers with a finite bandwidth to resonantly
excite all the above transitions near the trap center. The pumping process
takes about $5\tau _{IR}\simeq 0.05s$ ($\tau _{IR}$ is the IR natural
lifetime) which is considerably shorter than a typical trapping lifetime
(for example CaH lifetime is about $0.6s$\cite{buffer}).

Even after rotational pumping, there is a small fraction of the molecules
remaining in higher vibrational and rotational states. These molecules are
not in the 4 cooling cycle state ($|g0\rangle ,$ $|g\pm \rangle $ and $%
|e\rangle $)$.$ Therefore, they will not be translationally cooled and can
be disregarded. Among these 4 states, only a single state $|g0\rangle $ is
populated while the rest are essentially empty within a rovibrational decay
lifetime. Thus, we still have a large fraction of buffer gas cooled thermal
molecules in a single internal state $|g0\rangle $ as the precursor for
translational cooling. When the trapping field is switched off, the laser
cooling starts.

\section{Estimation of Cooling Parameters}

We show that the propose cooling scheme can be realized in practice. Based
on the results of our theoretical studies, we use realistic physical
parameters to briefly estimate the cooling time for each step. It is
essential to have a short total cooling time for unconfined cooling so that
the molecules are cooled before they fly apart beyond the laser cooling
region. Once translationally cooled, the molecules can be easily manipulated
and confined in a simple and shallow trap.

\subsection{\protect\bigskip Velocity selection}

The velocity selected momentum width can be arbitrarily narrow using long
duration Raman $\pi -$pulses. For practical purpose, it is essential to
consider an optimum pulse duration for this step. Since the selected
population is shifted by discrete number of $\hbar (k_{-}+k_{+})$ $\approx
2\hbar k$ in the deceleration step, it is most efficient to have the
selected width of $\Delta P_{vs}\approx 2\hbar k,$ corresponding to 4 times
the recoil temperature, which is $T_{vs}=\frac{(2\hbar k)^{2}}{2Mk_{B}}%
\approx 4\mu K$ \ for optical wavelength of $\lambda =300nm$ and molecular
mass of $M=100a.m.u.$ For Blackman pulse shape \cite{Harris}, the required $%
\pi -$pulses duration is estimated to be $\tau _{1}\approx \frac{15M}{%
k\Delta P_{vs}}=\frac{15}{4\omega _{r}}\approx 27\mu s,$ which is longer
than the natural lifetime $(\Gamma _{e}\approx 10^{7}s^{-1})$ of state $%
|e\rangle $ and the vibrational period of a molecule\cite{APLIP}$.$ Since
the sweep of the Blackman pulses is not a rapid passage, the excited state
will be appreciably populated unless the optical detuning $\delta $ is large
enough, but smaller than the rotational level spacing (around $%
10^{10}s^{-1}) $ to avoid excitations to untargeted rotational states. We
use $\delta =-50\Gamma \approx 5\times 10^{8}s^{-1}.$ The corresponding Rabi
frequency amplitude is $\Omega _{o}=\sqrt{-\delta \pi /(\tau _{1}0.61)}%
\approx 9.8\times 10^{6}s^{-1}$\cite{derive}.

\bigskip\ For a molecular gas of mass $100a.m.u.$ and initial temperature $%
T_{i}=1K,$ the momentum width is The maximum mean momentum is taken to be $%
P_{\max }=\sqrt{2Mk_{B}T_{i}}\approx 1000\hbar k.$ If each velocity
selection step selects a momentum width of $\Delta P_{vs}=2\hbar k,$ the
total number of velocity selection steps which is the number of cooling
cycle, $N_{\max }=2P_{\max }/\Delta P_{vs}\approx 1000,$ The total time
required for velocity selection in the cooling process is $t_{1}=N_{\max
}\tau _{1}\approx 27ms.$

\subsection{\protect\bigskip Deceleration}

\bigskip For complete population transfer between states $|g+\rangle $ and $%
|g-\rangle $ in each STIRAP inversion, in addition to counterintuitive
sequence and overlapping pulses, the pulses duration $\tau _{2}$ must meet
the adiabatic criteria \cite{revstirap}. Our simulations which take into
account spontaneous emissions using master equations give a criteria of $%
\tau _{2}\Omega _{o}\gtrsim 50$ for $99.95\%$ transfer efficiency with zero
detuning. With an efficiency of $\epsilon =99.95\%$ for each STIRAP
inversion, the minimum deceleration efficiency corresponds to the maximum
number of inversions $N_{\max }=1000$ for molecules with mean $P_{N_{\max }}$%
, which is $\epsilon _{N}=\epsilon ^{N_{\max }}$ $=60\%$.

In one-photon resonant condition to intermediate state $|e\rangle
=|X,n=0,J=0,M_{J}=0\rangle $, the finite spectral width $\Delta \omega $ of
the pulses must be much smaller than rotational level spacing $\omega _{rot}$
to avoid off-resonant coupling to additional intermediate states which may
lead to imperfect transfer efficiency; for example, $\Delta \omega \approx
1/\tau _{2}<<\omega _{rot}\approx 10^{10}s^{-1}.$ Thus, one can safely use
the typical \cite{revstirap} nanosecond STIRAP pulses and have each
inversion duration of $\tau _{2}=10^{-8}s$ corresponding to the spectral
width of $\Delta \omega \approx 10^{8}s^{-1}$ and a moderate peak Rabi
frequency $\Omega _{o}\approx 5\times 10^{9}s^{-1}$.

The number of STIRAP inversions required to decelerate a velocity selected
slice with mean momentum $P_{N}$ is $N\approx P_{N}/2\hbar k$ and the
corresponding deceleration time is $t_{N}=N\tau _{2}.$ The total
deceleration time in the cooling process is estimated as

\begin{equation}
t_{2}=\sum\limits_{N=1}^{N\max }t_{N}=\sum\limits_{N=1}^{N\max }N\tau _{2}=%
\frac{1}{2}N_{\max }(N_{\max }+1)\tau _{2}\simeq 5ms
\end{equation}

Despite the large number of STIRAP inversions, the total cooling time in
step 2 is still less than the total time for velocity selection.

\subsection{\protect\bigskip Accumulation}

In practice, the decelerated molecules in $|g+\rangle $ or $|g-\rangle $ can
be driven resonantly by c.w. laser to an excited electronic state for
flourescence decay to many ground rovibrational levels within $\tau
_{3}=10/\Gamma _{e}\approx 1\mu s.$ The total accumulation time is $%
t_{3}\simeq N_{\max }\tau _{3}\approx 1ms.$ Thus, the total time for the
whole cooling process is $t_{cool}=t_{1}+t_{2}+t_{3}\approx 33ms.$ The
cooling can be done simultaneously on the positive and negative sides of the
momentum distribution. If the velocity selection and deceleration start from
the fastest molecules to the slowest, the drift distance for molecules with
momentum $P_{N}$ is estimated as $L_{N}=N2\hbar k\{\frac{1}{2}(N_{\max
}-N)(N_{\max }-N+1)\tau _{2}+(N_{\max }-N)(\tau _{1}+\tau _{3})\}$ which
gives a maximum drift distance $20cm$ for $N=475.$ The maximum drift
distance can be reduced by optimizing the order of velocity selection on
certain velocity class.

\section{Modelling of the Cooling Effect}

\subsection{\protect\bigskip Initial state and density matrices}

From the practical descriptions in Section III, we assume that initially\
the internal state is pure and uncorrelated to the center of mass momentum
state. Only the internal state $|g0\rangle $ is occupied. The center of mass
subsystem is in the thermal {\em Gibbs state }\cite{linblad83} and the
initial total density operator is written as

\begin{equation}
\hat{\rho}_{tot}(0)=\hat{\rho}_{I}(0)\otimes \hat{\rho}_{cm}(0)=|g0\rangle
\langle g0|\otimes \sum\limits_{P}W(P)|P\rangle \langle P|
\end{equation}
where $W(P)$ $=\frac{1}{Z}e^{-P^{2}/\sigma ^{2}}$ is the statistical weight, 
$Z=\sum\limits_{P}e^{-P^{2}/\sigma ^{2}}$ is the partition function$\ $and $%
\sigma =\sqrt{2Mk_{B}T}$ is the initial momentum width with effective
temperature $T$. On one hand the gas is coupled to lasers, while on the
other hand it forms an open system whereby it can exchange entropy with the
outside environment.

The quantum dynamical evolution of the molecules in each cooling step and
throughout the cooling process can be modeled analytically by using the
density matrix elements. This enables convenient evaluation of the internal
and center of mass subsystems probability distributions and the
corresponding entropies, which allow us to work out the dynamics and the
cooling efficiency. The density matrices of the total system, internal
subsystem and center of mass subsystem in internal basis $\{|a\rangle \}$
and center of mass momentum basis $\{|P\rangle \},$ are defined respectively
as

\begin{eqnarray}
\hat{\rho}_{tot}(t) &\doteq &\sum\limits_{a,b}\sum\limits_{P,P^{\prime
}}\rho _{ab}(P,P^{\prime },t)|a,P\rangle \langle b,P^{\prime }| \\
\hat{\rho}_{I}(t) &\doteq &Tr_{cm}\{\hat{\rho}_{tot}(t)\}=\sum%
\limits_{a,b}|a\rangle \langle b|\sum\limits_{P}\rho _{ab}(P,P,t) \\
\hat{\rho}_{cm}(t) &\doteq &Tr_{I}\{\hat{\rho}_{tot}(t)\}=\sum\limits_{P,P^{%
\prime }}|P\rangle \langle P^{\prime }|\sum\limits_{a}\rho _{aa}(P,P^{\prime
},t)
\end{eqnarray}
where $\rho _{ab}(P,P^{\prime },t)\doteq \langle a,P|\hat{\rho}%
_{tot}(t)|b,P^{\prime }\rangle $ are the total system density matrix
elements, $a,b\in \{e,g0,g\pm ,d,\{acc_{j}\}\}$ the internal states and $%
\{P,P^{\prime }\}$ the center of mass momenta along $\sigma \pm $ polarized
lasers axis (defined as $z$-axis).

The time evolution dynamics of the total system in step 1 and step 2 are
formally described by the 3-level Bloch equations\cite{aspect89}. In step 1
the total system states are $|g0,P\rangle ,|e,P\rangle $ and $|g+,P-\hbar
k\rangle .$ A STIRAP inversion from $|g+\rangle $ to $|g-\rangle $ in step 2
involves the momentum family of states $|g+,P+\hbar k\rangle ,|e,P\rangle $
and $|g-,P-\hbar k\rangle .$ Due to the reversal of the pulses during
subsequent inversion from $|g-\rangle $ to $|g+\rangle ,$ the family of
states becomes $|g-,P-\hbar k\rangle ,|e,P-2\hbar k\rangle $ and $%
|g+,P-3\hbar k\rangle .$ With the high populations transfer efficiency of
STIRAP, the center of mass coherences in the populations, for example $%
\langle g+,P-\hbar k|\hat{\rho}_{tot}(t)|g+,P-3\hbar k\rangle ,$ are
negligibly small. We start with populations in the internal state $%
|g0\rangle $ and the center of mass Gibbs state with no coherences in
momentum basis. So, the only non vanishing matrix elements are $\rho
_{00}(P,P,0)$. The processes of a velocity selection, a STIRAP inversion and
a spontaneous emission do not create center of mass coherences. Therefore,
the populations remain diagonal in momentum basis throughout the cooling
process: $\rho _{aa}(P,P^{\prime },t)=\rho _{aa}(P,P,t)\delta _{P,P^{\prime
}}.$

On the other hand, transitions between different internal states are
accompanied by momentum recoil. Thus, the matrix elements between different
internal states$(a\neq b)$ are non vanishing only for the center of mass
coherences $P\neq P^{\prime }:$ $\rho _{ab}(P,P^{\prime },t)\neq 0$ and $%
\rho _{ab}(P,P,t)=0.$ Accordingly, Eqs. (3) and (4) reduce to

\begin{eqnarray}
\hat{\rho}_{I}(t) &=&\sum\limits_{a}C_{a}(t)|a\rangle \langle a| \\
\hat{\rho}_{cm}(t) &=&\sum\limits_{P}f(P,t)|P\rangle \langle P|
\end{eqnarray}
where the internal and center of mass probability distributions $C_{a}(t)$
and $f(P,t)$ are respectively defined as

\begin{eqnarray}
C_{a}(t) &\doteq &\sum\limits_{P}\rho _{aa}(P,P,t) \\
f(P,t) &\doteq &\sum\limits_{a}\rho _{aa}(P,P,t)
\end{eqnarray}
Thus, we only need to evaluate the diagonal density matrix elements $\rho
_{aa}(P,P,t)$ in order to calculate the subsystem density matrices and the
corresponding entropies.

\section{Population dynamics}

{\em Step 1 of N-th cooling cycle: }To simplify the mathematics, we can
model the velocity selected momentum distribution by a Gaussian function of
a narrow width $\sigma _{vsel}(<<\sigma ).$ During step 1 of the $N$-th
cooling cycle (subscript N:1), only the populations in the states $%
|g0\rangle $ and $|g+\rangle $ change with time and can be written
analytically as the following:

\begin{eqnarray}
\rho _{00}(P,P,t_{1})_{N:1}
&=&W(P)-\sum\limits_{j=1}^{N-1}V_{j}(P)-V_{N}(P)h_{1}(t_{1}) \\
\rho _{++}(P,P,t_{1})_{N:1} &=&h_{1}(t_{1})W(P_{N})e^{-(P+\alpha \hbar
k_{N,+}-P_{N})^{2}/\sigma _{vsel}^{2}}
\end{eqnarray}
where $0$ $\leq t_{1}\leq \tau _{1}$. $\tau _{1}$ is the duration of the
population transfer in step 1. The function $h_{1}(t_{1})=sin^{2}\pi t/2\tau
_{1}$ describes the time evolution of the populations. $V_{j}(P)\doteq
W(P_{j})e^{-(P-P_{j})^{2}/\sigma _{vsel}^{2}}$ is the velocity selected
distribution in the $j$-th$(j=1...N)$ cooling cycle with mean momentum $%
P_{j},$ momentum width $\sigma _{vsel}$ and mean population $W(P_{j})\doteq 
\frac{1}{Z}e^{-P_{j}^{2}/\sigma ^{2}}.$ $k_{N+}$ is the $\sigma ^{+}$ laser
wavevector of the $N$-th cooling cycle and $\alpha =sign(P_{j})$ $\det $er$%
\min $es the direction of the $\sigma ^{+}$ laser to provide decelerating
momentum kick.

The first term in Eq. (9) describes the initial populations distribution.
The second term corresponds to the depleted populations after $N-1$ cooling
cycles. The third term describes the transient population depletion in the $%
N $-th cooling cycle. In the end of step 1, the populations become

\begin{eqnarray}
\rho _{00}(P,P,\tau _{1})_{N:1} &=&W(P)-\sum\limits_{j=1}^{N}V_{j}(P) \\
\rho _{++}(P,P,\tau _{1})_{N:1} &=&W(P_{N})e^{-(P+\alpha \hbar
k_{N,+}-P_{N})^{2}/\sigma _{vsel}^{2}}
\end{eqnarray}

{\em Step 2 of N-th cooling cycle: }There are many substeps within step 2.
Only the populations in states $|g+\rangle $ and $|g-\rangle $ change with
time during the $n$-th substep of step 2 in the $N$-th cooling cycle,
according to

\begin{eqnarray}
\rho _{aa}(P,P,t_{2})_{N:2(n)} &=&h_{2}(t_{2})W(P_{N})e^{-(P+\alpha \hbar
k_{N,+}+\alpha \Lambda _{n}-P_{N})^{2}/\sigma _{vsel}^{2}} \\
\rho _{bb}(P,P,t_{2})_{N:2(n)} &=&(1-h_{2}(t_{2}))W(P_{N})e^{-(P+\alpha
\hbar k_{N,+}+\alpha \Lambda _{n-1}-P_{N})^{2}/\sigma _{vsel}^{2}} \\
\rho _{++}(P,P,0)_{N:2(1)} &=&\rho _{++}(P,P,\tau _{1})_{N:1}
\end{eqnarray}
where $a=-,$ $b=+$ for odd $n$(inversion from $|g+\rangle $ to $|g-\rangle $%
) and $a=+,$ $b=-$ for even $n.$ The time interval for every substep is $0$ $%
\leq t_{2}\leq \tau _{2}$, each with the same duration $\tau _{2}$ for
complete inversion. $h_{2}(t_{2})=sin^{2}\pi t/2\tau _{2}$ describes the
time evolution of the populations and $\Lambda _{n}\doteq
\sum\limits_{j=1}^{n}\hbar (k_{j+}+k_{j-})\approx $ $2n\hbar k$ is the total
amount of momentum transfer provided by $n$ pairs of STIRAP pulses. The
precise values of $k_{N+},$ $k_{j+}$ and $k_{j-}$ are obtained via the
two-photon Raman resonance condition. In the numerical simulations they are
simply taken to be approximately equal to $k$ without affecting the validity
of the results.

The momentum width $\sigma _{vsel}$ is unchanged throughout step 2. The
number of deceleration cycles $n_{\max }$ required to reduce the mean
momentum from $P_{N}$ to zero can be predicted through $\alpha \hbar
k_{N,+}+\alpha \Lambda _{n}-P_{N}\approx \alpha \hbar k(1+2n_{\max
})-P_{N}\approx 0,$ which gives

\begin{equation}
1\leq n\leq (n_{\max }=\frac{P_{N}}{2\alpha \hbar k}-\frac{1}{2}).
\end{equation}
In the end of step 2, the populations become

\begin{equation}
\rho _{aa}(P,P,\tau _{2})_{N:2(n_{\max })}\approx W(P_{N})e^{-P^{2}/\sigma
_{vsel}^{2}}
\end{equation}

{\em Step 3 of N-th cooling cycle: }After step 2, if the cooled populations
from the Raman states $|g\pm \rangle $ are transferred rapidly to the
decaying state $|d\rangle $ with a short $\pi $-pulse, the process is
essentially unitary due to negligible spontaneous emissions from $|d\rangle $%
. Furthermore, if the laser is orthogonal to the $z$-direction, no momentum
kick is imparted in $z$-direction. This enables the replacement $\rho
_{dd}(P,P,0)_{N:3}$ $\approx \rho _{aa}(P,P,\tau _{2})_{N:2(n_{\max })}$ as
initial population, setting $\rho _{\pm \pm }(P,P,t_{3})_{N:3}=0$ and
disregarding the dynamics of the $\pi -pulse.$ The population in $|d\rangle $
decay to many vibrational-rotational levels in the ground electronic state.
Assuming that only a small branching ratio of the population from $|d\rangle 
$ decays back to $|g\pm \rangle ,$ the time evolution of the populations in
step 3 is simply due to the spontaneous decays from $|d\rangle $ to $M$
number of accumulation states $\{|acc_{j}\rangle ,j=1...M\}$ with the
corresponding decay rates $\Gamma _{j}.$ This is given by 
\begin{eqnarray}
\rho _{dd}(P,P,t_{3})_{N:3} &=&\sum\limits_{j=1}^{M}F_{j}e^{-\Gamma
_{j}t_{3}}W(P_{N})e^{-P{}^{2}/\sigma _{vsel}^{2}} \\
\rho _{acc_{j}}(P,P,t_{3})_{N:3} &=&F_{j}\frac{\sigma _{vsel}}{\sigma _{acc}}%
e^{-P{}^{2}/\sigma
_{acc}^{2}}\{\sum\limits_{j=1}^{N-1}W(P_{j})+W(P_{N})(1-e^{-\Gamma
_{j}t_{3}})\}
\end{eqnarray}
where $0$ $\leq t_{3}\leq \tau _{3}$ is the time interval for step 3, $\tau
_{3}$ is the duration for $|d\rangle $ to empty its populations to
essentially zero (taken as $10/\Gamma _{j})$ and $F_{j}$ is the
Franck-Condon factor for level $|acc_{j}\rangle $ with $\sum%
\limits_{j=1}^{M}F_{j}=1.$

The first term in the curly bracket of Eq. (19) corresponds to the
cumulative populations from the previous $N-1$ cooling cycles. It is equal
to the depleted populations from $|g0\rangle $ while the second term
describes the transient increase in populations. The accumulated momentum
width $\sigma _{acc}$ is slightly larger than $\sigma _{vsel}$ due to the
small momentum spread (about $0.5\hbar k_{a}$) from a single spontaneous
emission. The normalization factor $\frac{\sigma _{vsel}}{\sigma _{acc}}$ in
Eq. (19) ensures that $Tr_{tot}\{\hat{\rho}_{tot}\}=1$ and can be estimated
by

\begin{equation}
\sigma _{acc}/\sigma _{vsel}\approx 1+0.5\hbar k_{a}/\sigma _{vsel}
\end{equation}
where $k_{a}=(E_{d}-E_{acc})/\hbar c.$

\section{Entropies}

The cooling effect is shown by the variation in the momentum probability
distribution and the variations of the total, internal and center of mass
von Neumann entropies\cite{neumann27}. From Eq. (1) the initial internal,
center of mass and total entropies are respectively:

\begin{eqnarray}
S_{I}(0) &=&0 \\
S_{cm}(0) &=&-k_{B}\sum\limits_{P}W(P)\ln W(P) \\
S_{tot}(0) &=&S_{cm}(0)
\end{eqnarray}
The expressions for internal and center of mass entropies are obtained from
Eqs. (5)-(8) using $S_{X}(t)=Tr_{X}\{\hat{\rho}_{X}(t)\ln \hat{\rho}_{X}\}$
where $X\in \{I,cm\},$ as follows:

\begin{eqnarray}
S_{I} &=&-k_{B}\sum\limits_{a}C_{a}\ln C_{a} \\
S_{cm} &=&-k_{B}\sum\limits_{P}f(P)\ln f(P)
\end{eqnarray}
Equations (24) and (25) are valid for all times in all cooling steps and
take a simple form because the reduced density matrices are diagonal
(Eqs.(5) and (6)). The center of mass probability distributions for step 1,
2 and 3 of the $N$-th cooling cycle are given by: 
\begin{eqnarray}
f(P,t_{1})_{N:1} &=&\sum\limits_{a=0,+}\rho
_{aa}(P,P,t_{1})_{N:1}+\sum\limits_{j}\rho _{acc_{j}}(P,P,\tau
_{3})_{(N-1):3} \\
f(P,t_{2})_{N:2(n)} &=&\rho _{00}(P,P,\tau
_{1})_{N:1}+\sum\limits_{a=+,-}\rho
_{aa}(P,P,t_{2})_{N:2(n)}+\sum\limits_{j}\rho _{acc_{j}}(P,P,\tau
_{3})_{(N-1):3} \\
f(P,t_{3})_{N:3} &=&\rho _{00}(P,P,\tau _{1})_{N:1}+\rho
_{dd}(P,P,t_{3})_{N:3}+\sum\limits_{j}\rho _{acc_{j}}(P,P,t_{3})_{N:3}
\end{eqnarray}
Using $\int_{-\infty }^{\infty }e^{-(P-a)^{2}/\sigma _{vsel}^{2}}dP=\sigma
_{vsel}\sqrt{\pi },$ we can show that $f(P)$ is normalized, $\int f(P)dP=1$.
Similarly, the internal probability distributions for step 1, 2 and 3 of the 
$N$-th cooling cycle are given by:

\begin{eqnarray}
C_{0,+}(t_{1})_{N:1} &=&\sum\limits_{P}\rho _{00,++}(P,P,t_{1})_{N:1} 
\nonumber \\
C_{acc_{j}}(t_{1})_{N:1} &=&\sum\limits_{P}\rho _{acc_{j}}(P,P,\tau
_{3})_{(N-1):3}
\end{eqnarray}
\begin{eqnarray}
C_{0}(t_{2})_{N:2(n)} &=&\sum\limits_{P}\rho _{00}(P,P,\tau _{1})_{N:1} 
\nonumber \\
C_{+,-}(t_{2})_{N:2(n)} &=&\sum\limits_{P}\rho _{++,--}(P,P,t_{2})_{N:2(n)} 
\nonumber \\
C_{acc_{j}}(t_{2})_{N:2(n)} &=&\sum\limits_{P}\rho _{acc_{j}}(P,P,\tau
_{3})_{(N-1):3}
\end{eqnarray}
\begin{eqnarray}
C_{0}(t_{3})_{N:3} &=&\sum\limits_{P}\rho _{00}(P,P,\tau _{1})_{N:1} 
\nonumber \\
C_{d}(t_{3})_{N:3} &=&\sum\limits_{P}\rho _{dd}(P,P,t_{3})_{N:3}  \nonumber
\\
C_{acc_{j}}(t_{3})_{N:3} &=&\sum\limits_{P}\rho _{acc_{j}}(P,P,t_{3})_{N:3}
\end{eqnarray}

The{\em \ total }density matrix in this step involves the basis states $%
|g0\rangle ,$ $|d\rangle $ and $\{|acc_{j}\rangle \}.$ It is diagonal since
these states are not coupled by laser. The evolution is characterized by
free spontaneous emission from $|d\rangle $ to $\{|acc_{j}\rangle \}.$Thus,
the time evolution of the total density matrix for step 3 can be written in
a completely diagonal form

\begin{equation}
\hat{\rho}_{tot}(t_{3})_{N:3}=\sum\limits_{P}\{\rho _{00}(P,P,\tau
_{1})_{N:1}|0,P\rangle \langle 0,P|+\sum\limits_{a=d,\{acc_{j}\}}\rho
_{aa}(P,P,t_{3})_{N:3}|a,P\rangle \langle a,P|\}
\end{equation}
with $Tr_{tot}\{\hat{\rho}_{tot,N}(t)\}=1.$

The total entropy changes only during incoherent process in step 3. Its time
evolution in the $N$-th cooling cycle is given by

\begin{equation}
S_{tot}(t_{3})_{N:3}=-k_{B}\sum\limits_{P}\{\rho _{00}(P,P,\tau
_{1})_{N:1}\ln \rho _{00}(P,P,\tau
_{1})_{N:1}+\sum\limits_{a=d,\{acc_{j}\}}\rho _{aa}(P,P,t_{3})_{N:3}\ln \rho
_{aa}(P,P,t_{3})_{N:3}\}
\end{equation}
Throughout the coherent interactions of step 1 and step 2 in the $N$-th
cooling cycle, the total entropy takes a constant value of $S_{tot}(\tau
_{3})_{(N-1):3}.$

Each cooling cycle is completed in step 3, during which radiation entropy $%
S_{R}(t)$ is created due to the random direction and many frequencies of the
spontaneously emitted photons as the cooled molecules decay to the many $%
|acc\rangle $ states. From the knowledge of the angular distribution of the
spontaneously emitted photons, we can write the radiation density matrix as

\begin{equation}
\hat{\rho}_{R}=\frac{1}{2\pi }\sum\limits_{j,{\bf \lambda }}\int_{0}^{2\pi
}\int_{0}^{\pi }F_{j}M_{j{\bf \lambda }}(\theta )|j,{\bf \lambda ,}\theta
,\phi \rangle \langle j,{\bf \lambda ,}\theta ,\phi |\sin \theta d\theta
d\phi
\end{equation}

where $|j,{\bf \lambda ,}\theta ,\phi \rangle $ is the state of the photon
emitted with frequency $\omega _{acc_{j}}=(E_{d}-E_{acc_{j}})/\hbar ,$
polarization $\lambda =0,\pm 1$, and in spherical coordinate angles $(\theta
,\phi )$ relative to the dipole axis. $M_{{\bf \lambda }}(\theta )=\frac{3}{8%
}(1+\cos ^{2}\theta )\delta _{{\bf \lambda },\pm 1}+\frac{3}{4}\sin
^{2}\theta \delta _{{\bf \lambda },0}$ corresponds to the probability of
emission at angle $\theta $ relative to the dipole axis and $F_{j}$ is the
Franck-Condon factor. The trace over all decay fequencies and angles gives $%
Tr_{j,{\bf \lambda ,}\theta ,\phi }\{\hat{\rho}_{R}\}=1.$ The radiation
entropy corresponding is easily obtained as

\begin{equation}
S_{R}(t)=-k_{B}\sum\limits_{j,{\bf \lambda }}\int_{0}^{\pi }R_{j{\bf \lambda 
}\theta }\ln R_{j{\bf \lambda }\theta }d\theta
\end{equation}

where $R_{j{\bf \lambda }\theta }=F_{j}M_{j{\bf \lambda }}(\theta )\sin
\theta .$

We note that the existence of many $|acc\rangle $ states during step 3 leads
larger radiation entropy due to the frequency dispersion, in addition to the
angular dispersion of the spontaneously emitted photons.

\section{Results and Discussions}

Based on the model discussed above, we simulate the essential features of
the cooling process. The typical variations of the internal populations
distribution for each step in a cooling cycle is shown in Figs. 3. In each
cooling cycle, a fraction of molecules with a narrow momentum width are
depleted from the state $|g0\rangle $ by velocity selection (Fig. 3a),
slowed towards zero mean momentum via repeated inversion by STIRAP (Fig. 3b)
and dumped irreversibly into the accumulation states (Fig. 3c).

We give the overall results of the cooling process first and discuss the
details of the cooling mechanism later. The cooling effect is shown by the
narrowing of the momentum probability distribution $f(P),$ which evolves
from a large momentum width (about $30\hbar k)$ into a narrow momentum width
(about $2\hbar k)$ with zero mean momentum (Fig. 6). The main interest of
this paper is the center of mass entropy $S_{cm}$ which is found to decrease
after the cooling process (Fig. 5a), in particular during velocity selection
and deceleration since these steps move a fraction of the populations closer
to zero momentum. This is the sign of cooling in the molecular translational
degree of freedom. The coherent processes of steps 1 and 2 contribute
substantially to the reduction of $S_{cm},$ from the maximum entropy of
Gibbs state to the final populations with narrow momentum width. Throughout
the cooling, the internal entropy $S_{I}$ increases(Fig. 5a). Within each
cooling cycle, the laser fields have created correlations between internal
and center of mass momentum states, thus transferring the entropy from the
center of mass to the internal subsystems by reducing the center of mass
state dispersion while increasing the internal state dispersion. Indefinite
repetition of the center of mass cooling effect is possible through the
irreversible process of step 3.

We proceed to elaborate on the detailed features of the entropy variations
in each cooling cycle with regard to the populations of the internal and
center of mass states. During velocity selection, the extraction of a narrow
slice of momentum from $|g0\rangle $ leads to distortion in the momentum
distribution. One $\hbar k$ of momentum kick on the velocity selected
population leads to a narrower overall momentum distribution and to the
reduction of $S_{cm}$(Fig. 4a). At the same time, $S_{I}$ increases(Fig. 4b)
because an additional internal state $|g+\rangle $ is increasingly populated.

In the deceleration step, $S_{cm}$ is reduced in each substep (Fig. 4a) by
repeated momentum kicks which push the selected populations closer to zero
mean momentum. Entanglement \cite{entangleSTIRAP} of the center of mass
momentum states in each STIRAP process increases the population dispersion
of the center of mass states. This explains the transient increase in $%
S_{cm},$ which is less than the overall reduction of $S_{cm}$ in one STIRAP
process. During the first STIRAP process in step 2, this transient increase
is not observed because it is quenched by the rapid reduction of $S_{cm}.$
The first STIRAP process reduces $S_{cm}$ more than the subsequent ones.
This quantitative difference is due to the nonlinear dependence of the
entropy on the momentum distribution. However, $S_{I}$ oscillates above a
constant value (Fig. 4b). Each oscillation corresponds to a substep, which
is an inversion process by STIRAP and can be explained by the creation of
adiabatic states through the entanglement of states $|g+\rangle $ and $%
|g-\rangle .$ The entanglement involves the population of an additional
internal state, thereby increasing $S_{I}.$ After the completion of a STIRAP
process, the entanglement disappears and $S_{I}$ falls back to the value
before the entanglement. Thus, the value of $S_{I}$ before and after step 2
are the same.

The total entropy $S_{tot}$ is constant throughout steps 1 and 2 but changes
only during step 3 (Fig. 4c). This shows that unitary evolution can reduce
the center of mass entropy despite the invariance of the total entropy. In
other words, coherent interaction alone is able to cool the center of mass
subsystem at the expense of heating the internal subsystem. However, the
cooling cannot be repeated unless an irreversible process is introduced as
part of the cooling cycle, which is step 3. In step 3, the slight increase
in $S_{cm}$ due to momentum spread from one spontaneous emission is smaller
than the decrease in $S_{cm}$ in steps 1 and 2 (Fig. 4a). This leads to the
overall reduction of $S_{cm}$ in each cooling cycle. In fact, it is the
coherent process of steps 1 and 2 instead of the spontaneous emission, which
substantially reduces $S_{cm}.$ This is an essential point, as it indicates
that {\em coherent process can lead to substantial change in the subsystem
entropy.} The important role of the single spontaneous emission is to {\em %
irreversibly accumulate} the translationally cooled populations{\em . }We
note that this accumulation role can also be realized if the cooled
populations in each cooling cycle are\ coherently transferred to different
internal $|acc\rangle $ states. For this, the number of $|acc\rangle $
states has to be more than the number of cooling cycles and multi-frequency
lasers resonant with all the $|acc\rangle $ states would be required. In
this case, the whole cooling process is coherent and {\em no spontaneous
emission} is required. Then, both translational and internal cooling cannot
be achieved simultaneously. Translational cooling must be accompanied by
internal heating, or vice versa. Thus, coherent cooling of translational
motion is not a conceptual impossibility but rather a practical one, at
least at the present moment.

The entropy variations for a {\em single }$|acc\rangle $ state and {\em %
multi }$|acc\rangle $ states are shown in Fig. 4b and Fig. 4c (we use 10 $%
|acc\rangle $ states with logarithmic Franck-Condon factors and a fixed
decay rate). With single $|acc\rangle $ state, $S_{I}$ reduces during step 3
of the second cooling cycle because the number of internally populated
states reduces from three populated states $(|g0\rangle ,|g\pm \rangle
,|acc\rangle )$ to two states $(|g0\rangle ,|acc\rangle )$(Fig. 4b). The
center of mass entropy $S_{cm}$ is unaffected by the number of $|acc\rangle $
states (Fig. 4a). However, the internal entropy $S_{I}$ increases with more $%
|acc\rangle $ states since the number of populated internal states increases
(Fig. 4b). It is interesting to observe that the increase in $S_{I}$ from
populating 10 internal states(in step 3) is still smaller than the increase
in $S_{I}$ from velocity selection(step 1). This is because entropy is not
simply a measure of the number of occupied states, but also depends on the
relative populations among the states.

The internal and center of mass entropies fluctuate throughout the cooling
process (Fig. 5a). As the momentum width of the populations shrinks towards
a narrow velocity selected width in the cooling process$,$ the $S_{cm}$
reduces from the maximum entropy state towards zero. The small residual
populations in the right and left wings of the narrow peak (Fig. 6) is due
to inefficient velocity selection with Gaussian momentum profile. In
practice, a more efficient pulse shape can be tailored to selectively
transfer the narrow populations as close to a square momentum profile as
possible. This would bring the final $S_{cm}$ closer to zero in the end of
the cooling process. On overall, $S_{I}$ increases in every cooling cycle
due to the increase of the populations in the many $|acc\rangle $ states,
thus changing from populating an internal state $(S_{I}=0)$ to populating
multi internal $|acc\rangle $ states.

The total entropy $S_{tot}$ before and after the cooling do not seem to
change a lot (Fig. 5b). In fact, $S_{tot}$ can increase or decrease due to
the coupling of the total system to the environment. The total, internal and
center of mass entropies satisfy the Araki-Lieb inequality\cite{arakilieb70}%
, $|S_{I}-$ $S_{cm}|\leq S_{tot}\leq S_{I}+$ $S_{cm},$ as shown in Fig. 4c
and 5b. Therefore, the cooling results are consistent with the general
properties of quantum entropy. The non-negligible index of correlation $%
I_{C}=S_{cm}+S_{I}-S_{tot}$\cite{linblad83} shows that the internal and
center of mass subsystems are correlated by the laser interactions
throughout the cooling process. However, after cooling the correlation is
not maximum since $S_{tot}$ is not zero and $S_{cm}\neq S_{I}.$

We note that the increase in the internal entropy is due to the specific
choice of the accumulation scheme (step 3), due to practical requirement for
faster cooling time. The amount of increase in the internal and radiation
entropies depend essentially on the number of $|acc\rangle $ states. The
resulting internal heating is entirely analogous to creation of radiation
entropy. The many internal states in molecules provide an additional entropy
sink, much like the radiation Fock states. During the cooling process, the
translational entropy is removed irreversibly by spontaneous emissions to
these entropy sinks.

If we had coherently accumulated the cooled molecules by transferring each
cooled fraction of molecules after each cooling cycle to distinct
rovibrational states instead of using spontaneous emission, less radiation
entropy will be created while the internal entropy increases dramatically.
This requires multi-frequency lasers, which may not be practical. On the
other hand, we can use a low rovibrational state as $|d\rangle $ state which
has less decay channels. Then, infrared (IR) spontaneous emission will be
the accumulation mechanism instead of optical spontaneous emission and the
radiation entropy would still increase but with less populated $|acc\rangle $
states and therefore less internal entropy. However, the long IR decay
lifetime leads to longer cooling time.

The model presented above does not consider further IR decays from the $%
|acc\rangle $ states to lower rovibrational states. Within about $0.1s$, the
molecules essentially end up in the ground vibrational state by multiple IR
decays. Thus, the vibrational heating occuring in the model is a {\em %
transient }effect. We have shown the model up to a stage where translational
cooling has been achieved to a large extend, which is the aim of our cooling
scheme. To summarize, the resulting internal heating is not a practical
concern in this work since our aim is translational cooling. Translationally
cooled molecules (even if internally hot) are invaluable assets in many
fundamental and applied sciences.

\section{Conclusions}

We have developed a laser cooling concept for translational motion of
neutral molecules based on the repetition of three steps: velocity
selection, deceleration and accumulation by a single spontaneous emission
per molecule. Each molecule need not undergo repeated spontaneous emissions
to be translationally cooled. Using realistic cooling parameters, we have
estimated the cooling time, which turns out to be practically feasible to
cool down thermal molecules of $1K$ to around recoil width. We have modelled
the cooling process and studied the changes in the entropies. Simulations of
the model have shown the reduction of the momentum width and the center of
mass entropy by the cooling process. The many internal states of molecules 
can serve as a useful entropy sink, in addition to the radiation entropy for
translational cooling. Analysis has shown that our scheme to translationally
cool molecules by lasers, can be realized in pinciple and in practice. This
cooling concept should open up a new possibility for translational laser
cooling of molecules.

\begin{center}
\bigskip {\large ACKNOWLEDGMENTS}
\end{center}

We wish to thank Achim Peters, Dennis Weise and Thomas Konrad for
discussions. Financial support by the Deutsche Forschungsgemeinschaft
(Forschergruppe Quantengase) is gratefully acknowledged.


\newpage 
\begin{figure}[t]
\caption{A rotational pumping scheme for initial state preparation}
\label{fig1}
\end{figure}

\newpage 
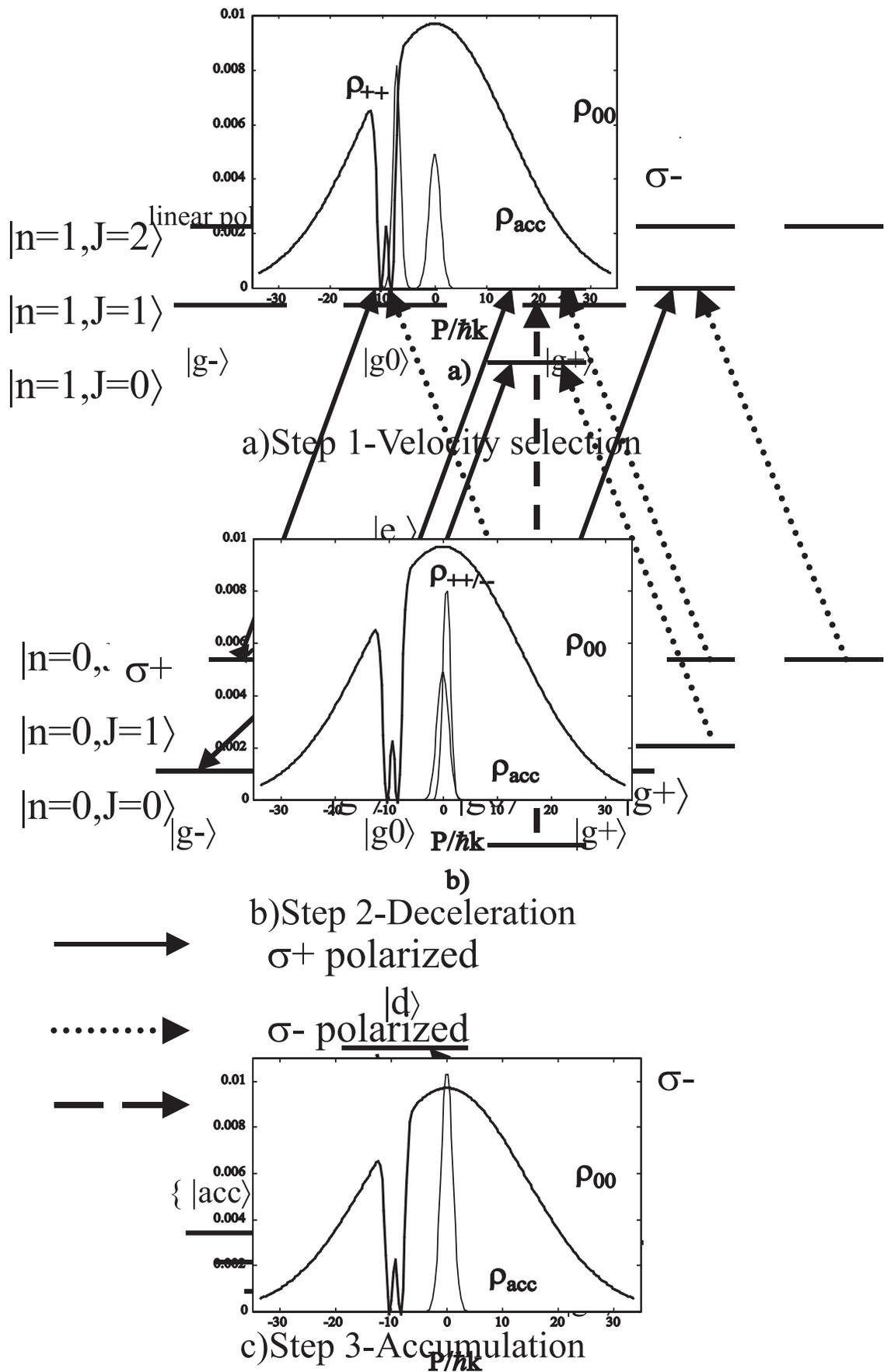
\begin{figure}[t]
\caption{Schematic diagrams for each cooling cycle composed of 3 sequential
steps: a)narrow velocity width selection, b)deceleration of selected
molecules and c)accumulation of decelerated and narrow momentum width
molecules by a single spontaneous emission to one of the many accumulation
states. }
\label{fig2}
\end{figure}

\newpage 
\begin{figure}[t]
\caption{Momentum distributions of the internal populations in a typical
cooling cycle after the steps of: a)velocity selection, b) deceleration and
c)accumulation.}
\label{fig3}
\end{figure}

\newpage 
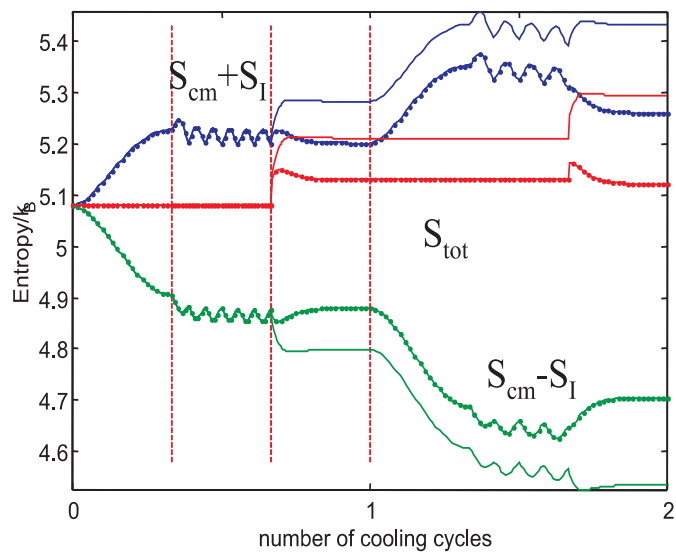
\begin{figure}[t]
\caption{Evolution of the a)center of mass entropy $S_{cm}$, b)internal
entropy $S_{I}$, and c)total entropy $S_{tot}$ as well as $S_{cm}\pm S_{I}$
throughout the first 2 cooling cycles with 1 accumulation state(line with
dots) and 10 accumulation states (solid line). The abscissa of each cooling
cycle is equally divided for the 3 cooling steps, each with different
timescale.}
\label{fig4}
\end{figure}

\newpage 
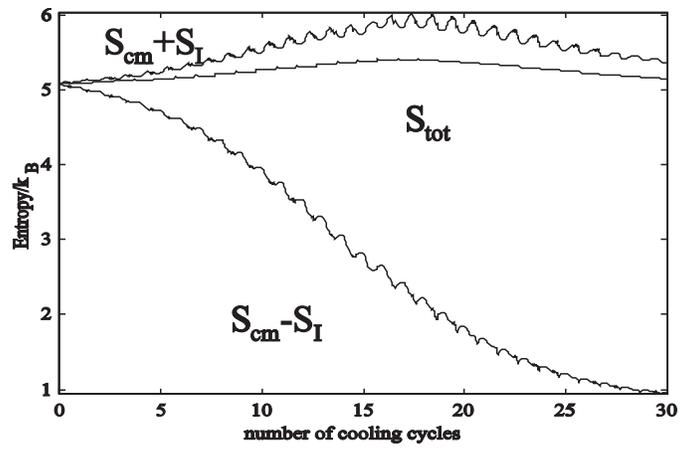
\begin{figure}[t]
\caption{Results after the whole cooling process with 10 accumulation states
for a)$S_{cm}$ and $S_{I}$, b)$S_{tot}$ and $S_{cm}\pm S_{I}$.}
\label{fig5}
\end{figure}

\newpage 
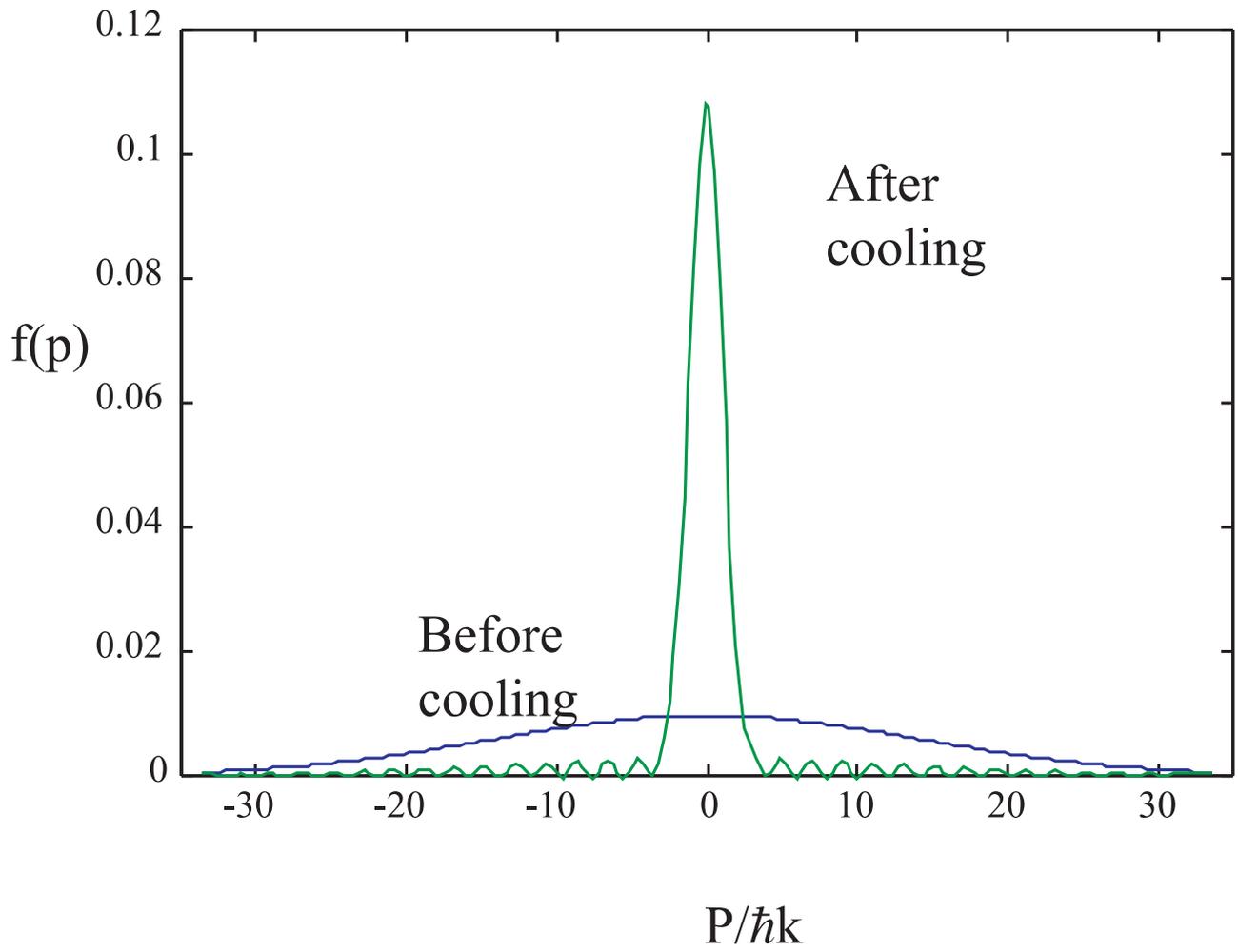
\begin{figure}[t]
\caption{Total momentum probability distribution $f(P)$ before and after the
cooling process. }
\label{fig6}
\end{figure}


\begin{references}
\bibitem{derive}  This equation is the Raman $\pi $-pulse criteria for
complete velocity selective population transfer derived from the effective
two level Bloch vector model under the two-photon Raman resonance condition
for Blackman $\pi -$pulse.
\end{references}
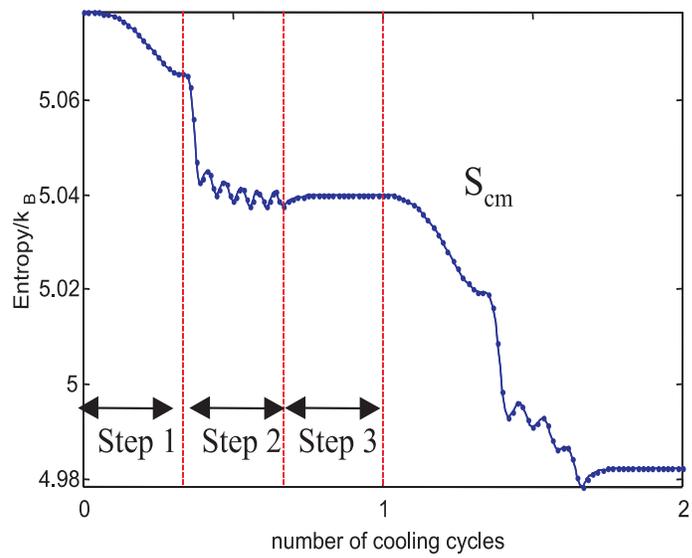
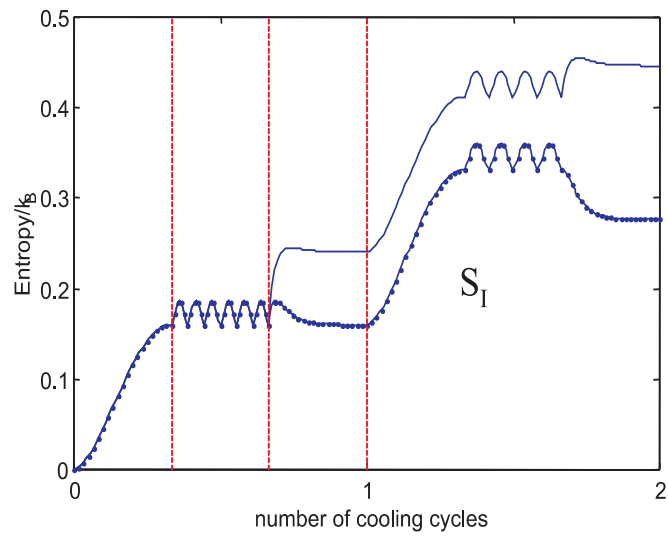
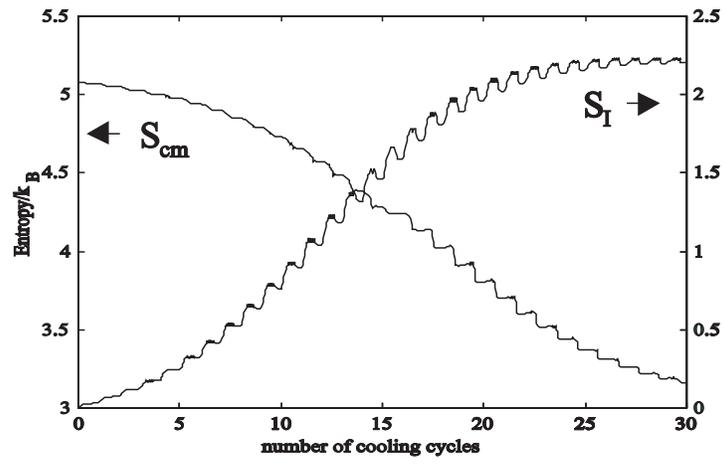
\end{document}